\documentclass[12pt,nofootinbib,a4paper]{revtex4-1}
\usepackage{amsmath,mathrsfs,amscd,graphicx,cancel}
\usepackage{multirow}
\usepackage{color}
\usepackage{rotating}
\usepackage{cancel}
\usepackage{subfig}
\usepackage{textcomp}
\usepackage[final]{pdfpages}
\usepackage{array}
\usepackage{float}
\usepackage{url}
\usepackage{textcomp}
\usepackage{amsfonts, latexsym, epsfig}
\usepackage{bm}
\usepackage{times}
\usepackage{epsfig}
\usepackage{amssymb}
\def\beq{\begin{equation}}
\def\eeq{\end{equation}}
\def\bea{\begin{eqnarray}}
\def\eea{\end{eqnarray}}

\begin{document}
\title{Squarkonium, diquarkonium and octetonium at the LHC and their di-photon decays }
\author{Ming-xing Luo}
\email{mingxingluo@zju.edu.cn}
\author{Kai Wang}
\email{wangkai1@zju.edu.cn}
\author{Tao Xu}
\email{taoxu@zju.edu.cn}
\author{Liangliang Zhang}
\email{zhngliang87@zju.edu.cn}
\author{Guohuai Zhu}
\email{zhugh@zju.edu.cn}
\affiliation{
Zhejiang Institute of Modern Physics and Department of Physics, Zhejiang University, Hangzhou, Zhejiang 310027, China
}
\begin{abstract}
Motivated by the recent di-photon excess by both ATLAS and CMS collaborations at the LHC, we systematically investigate the production and di-photon decay of onia formed by pair of all possible color exotic scalars in minimal extension. When such scalar massive meta-stable colored and charged (MMCC) particles are produced in pair near threshold, $\eta$ onium can be formed and decay into di-photon through annihilation as $pp\to \eta \to \gamma\gamma$. Squarkonium is formed by meta-stable squarks in supersymmetric models such as stoponium. Diquarkonium is formed by meta-stable color sextet diquarks which may be realized in the Pati-Salam model. Octetonium is formed by color octet scalars bosons as in the Manohar-Wise model. Stoponium prediction is much smaller than the required signal to account for the di-photon excess. Due to the enhancement factor from color and electric charge, predictions of diquarkonium and octetonium are of $\cal O$(10~fb) which are significantly greater than the stoponium prediction. Since the color enhancement also results in large production at the colliders, such light color exotic states of $\cal O$(375~GeV) suffer from severe  direct search constraints. On the other hand, if their dominant decay mode involve top quark, they may be buried in the $t\bar{t}$ plus jets samples and can potentially be searched via $t+j$ resonance. 
\end{abstract}
\maketitle

\section{Introduction}

Recently both the ATLAS and the CMS collaborations have reported an excess of $\gamma\gamma$ events  at the LHC Run-$2$ with a center-of-mass energy $\sqrt{s}=13$~TeV \cite{atlas,CMS:2015dxe}. This excess emerges with a bump in the di-photon invariant mass spectrum around $750$~GeV over the predicted continuous falling background. The ATLAS collaboration used $3.2$ fb$^{-1}$ of data and the local (global) significance is $3.6(2.0)\sigma$. The CMS collaboration observed a local (global) significance as $2.6(\text{less~than}1.2)\sigma$ with $2.6$ fb$^{-1}$ of data. The observed events have quite similar kinematic properties with the background events in this spectrum region. Even though a clearer picture requires more data, it is still worth studying the possible hint from this excess.
Considering the current volume of data sample, a rough estimation of the cross-section $\sigma(pp\rightarrow X \rightarrow \gamma\gamma)$ is of a few fb.

Landau-Yang theorem excludes the di-photon resonance from being vector boson and only spin-$0$ or spin-$2$ resonance can be viable candidate. $KK$-graviton predicts similar order of di-lepton decay which may suffer from direct constrain \cite{Randall:2008xg}. If the excess arises from a spin-$0$ resonance, candidate can be realized among fundamental scalar, a pion-like $\pi^{0}$ or an $\eta$-like resonance. Direct searches via $W^{+}W^{-}$ or $ZZ$ put stringent bound over the heavy Higgs-like scalar \cite{Aad:2015kna} and a CP-odd Higgs $A$ is then more plausible. Minimal model with CP-odd Higgs suffers from huge suppression of decay branching fraction because of decrease in $\gamma\gamma$ partial width due to lack of $W$-loop contribution and  enhancement of total width from on-shell $t\bar{t}$ decay. Many papers have been published to discuss the implications of the di-photon resonance \cite{Backovic:2015fnp,Agrawal:2015dbf} which cover proposals as CP-odd Higgs with exotic quark, sneutrino with $R$-parity violation, pseduo-scalar from chiral symmetry breaking of new strong dynamics.

A proton-proton collider like LHC is typically a Quantum Chromodynamics (QCD) machine where new physics in the strong interaction sector will appear in the early stages of operation even with limited luminosity. The recent di-photon anomaly can well be
a signal as onia formed by massive meta-stable colored and charged (MMCC) states. For a MMCC state $Q$, if the total width $\Gamma_{Q} <\Lambda_{\rm QCD}$, the particle $Q$ could form a hadronic bound state long before it decays.  Since $Q$ is colored, it can be produced in pair as $Q\bar{Q}$ at the LHC and they will form hadronic states as for instance $Q\bar{q}$ with light quarks from vacuum. In addition, near the threshold, they can form hadronic bound state $\eta(Q\bar{Q})$ as scalar onium. The onium $\eta_{Q}$ is produced through gluon fusion and decay into photon pair,
\beq
pp \to \eta_{Q} \to \gamma\gamma~.
\eeq
On the other hand, if $Q$ is fermionic, the hadronic bound states include both scalar and vector states in analogy to $\eta_{c}$ and $J/\psi$ case \cite{Agrawal:2015dbf}. The vector state will not only decay into di-photon but also decay into di-lepton which suffers from severe constraint of direct searches at LHC. However the $\eta$-like state can be produced via gluon fusion $gg\to \eta$ while the $J/\psi$-like state can only be produced through $q\bar{q}$ annihilation, therefore the production of the vector state is significantly lower than that of the scalar state at the LHC. Taking this into account, it requires more study on the LHC constraints on vector bound state of $Q\bar{Q}$ for fermionic $Q$. In the following we discuss only scalar MMCC cases.

We investigate three categories of such scalar MMCC to illustrate the
feature. The next section, we focus on stoponium resonance from light stop of compressed supersymmetry.
Then we discuss the  other two color exotics, color sextet scalar as diquark 
and the color octet scalars and their constraints.
We then discuss the di-photon prediction of diquarkonium formed by long-lived diquark/anti-diquark pair and the octectonium formed by color-octet scalars. We then conclude in the final section.

\section{Stoponium}
A first well-known example of such scalar MMCC state is a stop in the minimal supersymmetric standard model (MSSM). When meta-stable stops are produced in pair at colliders, stoponium can be formed near threshold  \cite{nojiri, martin, Kim:2014yaa} and stoponium can decay into $gg$, $ZZ$ $\gamma\gamma$ {\it etc.} through annihilation. Here we briefly summarize the di-photon signal from stoponium at the LHC which was studied in \cite{martin,Kim:2014yaa}.

As a consequence of large top Yukawa, third generation squarks are typically the lightest among
the sfermion spectrum and can even be the next lightest supersymmetric particle (NLSP). In compressed SUSY model,
the mass difference between stop and lightest neutralino,
$\Delta m= m_{\tilde{t}_{1}}- m_{\tilde{\chi}^{0}_{1}}$,
can be of a few GeV. Therefore, stop can only decay through
loop or four-body
\beq
\tilde{t}_{1}\to c\tilde{\chi}^{0}_{1}, ~~\tilde{t}_{1}\to b \ell\nu \tilde{\chi}^{0}_{1}~.
\eeq
The total width of stop is only of $\cal O$(KeV) which is less than $\Lambda_{QCD}$. When being produced, such stop will form hadronic state as
$R$-hadron, which is the composite colorless state of colored metastable superparticles with light quarks. However, if stop decay within the
tracking system, most of the $R$-hadron searches do not apply.
In addition, the stop decay is largely into neutralino with extremely soft objects like $D$-meson. The pair production signal is mostly
large missing transverse energy $\cancel{E}_{T}$. The only bound then comes from
$pp\to j+ \tilde{t}_{1} \tilde{t}_{1}^{*}$ and the signal is mono-jet plus
$\cancel{E}_{T}$. Without any kinematic handle, bound on such final state is rather weak. Dark matter annihilation is mainly through $\tilde{\chi}^{0}_{1}\tilde{\chi}^{0}_{1}\to t\bar{t}$ but may also have $\tilde{\chi}^{0}_{1}  \tilde{t}_{1}$ co-annihilation and it's not difficult to obtain the viable parameter region with correct relic density \cite{martin}.

On the other hand, pair production of stop does not necessarily lead to large $\cancel{E}_{T}$ which is common signature for all $R$-parity conserving MSSM. Stop production near threshold can also form hadronic bound state as stoponium $\eta_{\tilde{t}}$ as a typical example of squarkonium \cite{martin}. The stoponium can decay into $gg$, $\gamma\gamma$, $W^{+}W^{-}$, $ZZ$, $hh$, $t\bar{t}$, $b\bar{b}$ and $\tilde{\chi}^{0}_{1}\tilde{\chi}^{0}_{1}$ by annihilation. The stoponium production rate can be estimated from Higgs production from gluon fusion
by scaling from ratio of partial width $\Gamma(\eta\to gg)/\Gamma(H\to gg)$. A comprehensive phenomenology study has been performed by \cite{martin}. However, at 14~TeV LHC, the predicted $\sigma(pp\to \eta_{\tilde{t}_{1}}\to \gamma\gamma)$ is about 0.05~fb for $m_{\eta}=750$~GeV which is much less than the required rate to account for the current di-photon excess. Detailed calculations for stoponium production and decays to $\gamma\gamma$ or $ZZ$ are also presented in \cite{Kim:2014yaa}. A recent lattice study on the origin of the stoponium wave function \cite{Kim:2015zqa} indicates that the stoponium production rate may be about $3.5$ times larger than the potential model calculation \cite{Hagiwara:1990sq} adopted in \cite{martin}, which is still far too small to accommodate the LHC di-photon excess.

\section{Scalar Color Exotics}

Since the stoponium prediction is much lower to account for the excess, one would need large enhancement in production or di-photon decay. In this session, we discuss the scalar color exotics. In minimal extension, scalar exotics decay into fermionic quark pairs. Under $SU(3)_C$, 
\bea
{3}\otimes {3} &=& {6}\oplus{\bar{3}}\nonumber\\
{3}\otimes \bar{3} &=& {8}\oplus{1}~.
\eea
Therefore, the scalar color exotics can only be sextet, anti-triplet and octet under $SU(3)_{C}$. The above color structure corresponds to the following Lorentz structure respectively as
\beq
\bar{\psi^{c}}\psi \Phi + \bar{\psi}\psi \phi
\eeq
where $\psi$ is a Dirac spinor. The first term violate the global fermion number $B+L$ while the second term corresponds to fermion number conservation. The $\Phi$ field can be identified as sextet or anti-triplet scalar while the $\phi$ field can be identified as singlet or octet. Sextet and anti-triplet scalar both couple to quark quark pairs and hence are called diquark. Taking $\psi=P_{L} \psi + P_{R} \psi $, it is straightforward to conclude that diquark $\Phi$ couples to the same chiral field while the field $\phi$ couples to spinor with different chirality. Under $SU(2)_{L}$, diquark $\Phi$ can be either triplet or singlet
while $\phi$ can only be doublet. 

The anti-triplet diquark sometimes can be identified as scalar quark in supersymmetric theory. 
If $R$-parity is violated, squark can decay into quark pair through superpotential 
$\epsilon_{\alpha\beta\gamma}u^{c}_{\alpha}d^{c}_{\beta}d^{c}_{\gamma}$ coupling.
On the other hand,  stoponium in the previous session already gives the maximal prediction among squarkonia states.

As we argued, color octet scalars must be $\phi_{8}(8,2,1/2)$ under the SM gauge symmetry with Yukawa coupling 
\beq
y \bar{Q}_{L} u_{R} \phi_{8}+h.c.~. 
\eeq
The octet scalars has been partially studied in Manohar-Wise model \cite{Manohar:2006ga}.

The color sextet scalar 
is a symmetric $2^{\rm nd}$ rank tensor of $SU(3)_{C}$. 
Diquarks couple to SM quarks as Fermion number violation $\psi^T C^{-1} \psi \phi$, where $\psi$ is a Dirac spinor and $\phi$ is the scalar diquark.  In $SU(3)_C\times SU(2)_L\times U(1)_Y$, the sextet diquark can be 
$\Delta_6$, a $SU(2)_L$ adjoint $(6,3,1/3)$ and $SU(2)_L$ singlets 
\beq
\Phi^{+4/3}_6: (6,1,4/3); ~~~~
\Phi^{-2/3}_6: (6,1,-2/3); ~~~~
\Phi^{+1/3}_6: (6,1,+1/3)~.
\eeq 
Sextet diquark can be identified as a color sextet scalar in Pati-Salam
model $SU(4)_{C}\times SU(2)_{R}\times SU(2)_{L}$. During the symmetry breaking of
\beq
SU(4)_{C}\times SU(2)_{R}\times SU(2)_{L} \to
SU(3)_{C}\times SU(2)_{R}\times SU(2)_{L} \times U(1)_{B-L}~,
\eeq
a $\bf 10$-dimensional symmetric second rank tensor under $SU(4)$ can be decompose as
\bea
(10,1,3)& = & (6,3,1,2/3)+ (3,3,1,-2/3)+ (1,3,1,-2)\nonumber\\
(10,3,1)& = & (6,1,3,2/3)+ (3,1,3,-2/3)+ (1,1,3,-2)
\label{triplet}
\eea
Even though the scale of $SU(2)_R\times SU(4)_C$ symmetry breaking
is around $10^{10}$ GeV, in a supersymmetric Pati-Salam model \cite{Chacko:1998td,willie}, light color sextet scalar (diquark) can be realized as a result of existence of accidental symmetries where the masses of color sextet scalar only arises through high dimension operators. All the diquark states are charged under electromagnetic interaction and hence, non-zero vacuum expectation value is strictly forbidden for diquarks.  Diquark carries non-zero $U(1)_{B}$ Baryon number while the $SU(2)$ triplet state in Eq. (\ref{triplet}) carries non-zero $U(1)_{L}$ Lepton number.
When $SU(2)$ triplet  acquire vacuum expectation value during symmetry breaking, Majorana neutrino mass naturally arises and Lepton number is violated by two units. Due to $U(1)_{B-L}$ gauge symmetry, the lepton number violation can also be converted into baryon number violation
\beq
\Delta L=2  \to \Delta B=2~.
\eeq
However,
such $B/L$ violation does not lead to proton decay which is $\Delta B=1$ and $\Delta L=1$ effect,  $\Delta B=2$ violation only leads to neutron-anti-neutron ($n-\bar{n}$) oscillation and the electroweak scale diquark is fully compatible with present
limits \cite{Chacko:1998td}. In addition, such diquark also helps in Post-Sphaleron baryogenesis \cite{babu}.

The diquark decay through Yukawa coupling 
\beq
 f_\Phi u^T_R C^{-1} u_R \Phi^\dagger_6~
\eeq
with the 
width as \cite{willie}
\beq
\Gamma_{ij} = {3\over 8\pi (1+\delta_{ij})} |f_{ij}|^2 M_{\Phi_6} \lambda^{1/2}(1,r^2_i,r^2_j)(1-r^2_i-r^2_j)
\eeq
where $\lambda(x,y,z)=(x-y-z)^2-4 y z$ and $r_i = m_i/M_{\Phi_6}$.
The Yukawa couplings are typically constrained by flavor physics. For instance, for the $SU(2)_{L}$ singlet with only coupling to the up-type quarks as $\Phi_{6} (6,1,4/3)$, the most stringent bounds on the couplings $f_{ij}$ come from $D^0-\bar{D}^0$ mixing, to which $\Phi_6$ would make a tree level contribution proportional to $f_{11}f_{22}/M^2_{\Phi_6}$. The off-diagonal coupling $f_{ij}$ will contribute
to flavor violation processes, for instance $D\to \pi\pi$ which is proportional to $f_{12}f_{11}/M^2_{\Phi_6}$. The current bounds
require that
\beq
f_{11} f_{22} \lesssim 10^{-6}; f_{11} f_{12} \lesssim 10^{-2},
\eeq
for $M_{\Phi_6}$ of a few hundred GeV to TeV mass range \cite{chen}. Less stringent constraint comes from one loop process as $c\to u\gamma$. To escape from the bound, the charm-related couplings should be negligible.
 In the region of $f\lesssim 10^{-3}$, $\Phi_{6}$ decay width is less than $\Lambda_{\rm QCD}$ \cite{willie} and $\Phi_{6}$ becomes meta-stable.

\section{Diquarkonium and Octetonium}

We first use the $\Phi^{+4/3}_{6}$ to illustrate the feature then scale down to other sextet diquarkonia.
Near threshold, $\Phi_{6}\Phi_{6}^{\dagger}$ can form diquarkonium $\eta_{\Phi}$ which would decay dominantly into the $gg$ channel. Since $\Phi^{+4/3}_{6}$ is a $SU(2)_{L}$ singlet, $\eta_{\Phi}$ decays also into $\gamma\gamma$, $Z\gamma$ and $ZZ$, with the latter two channels suppressed by $\tan^2 \theta_W$ and $\tan^4 \theta_W$ in comparison to the di-photon channel.  Therefore
 $\text{Br}(\eta_{\Phi}\to \gamma\gamma) \simeq \Gamma(\eta_{\Phi}\to \gamma\gamma)/\Gamma(\eta_{\Phi}\to gg)$.

Since the gluon fusion production of such diquarkonium is proportional to the partial width $\Gamma(\eta_{\Phi}\to gg)$, at leading order, the production rate is
\beq
{\sigma(pp\to \eta_{\Phi})\over \sigma(pp\to H)}= {\Gamma(\eta_{\Phi}\to gg)\over \Gamma(H\to gg)}~.
\eeq
We find the partial decay width of $\eta_\Phi \to gg$ at leading order as
\beq
\Gamma(\eta_\Phi \to gg)={50 \alpha_S^2 \over 3  } {\mid R(0)\mid^{2} \over m^{2}_{\eta_\Phi}}~,
\eeq
in which the numerical factor is $25/2$ times larger than that of $\Gamma(\eta_{\tilde{t}} \to gg)$. Here $R(0)$ is the radial wave function at the origin.
The large color-factor enhancement in the above digluon decay comes from the relation $Tr(T^A T^B)=\delta^{AB}/2$ in the fundamental representation
while $Tr(T^A T^B)=5\delta^{AB}/2$ in the sextet representation of $SU(3)_C$. The factor $1/2$ arises from the normalization of the wave function at color space, which is $1/\sqrt{6}$ for diquarkonium and $1/\sqrt{3}$ for stoponium.

Similarly, the di-photon decay width at leading order is
\beq
\Gamma(\eta_\Phi \to \gamma\gamma)={1024 \alpha^2 \over 27 } {\mid R(0)\mid^{2} \over m^{2}_{\eta_\Phi}}~,
\eeq
in which the numerical factor is $32$ times larger than that of $\Gamma(\eta_{\tilde{t}} \to \gamma\gamma)$.
Here part of the enhancement (factor of $16$) comes from the diquark electric charge which is twice larger than that of stop, another factor of 2
enhancement is due to the fact that there are six colors for diquarks instead of three colors for stops.
It is expected that $R(0)$ of diquarkonium should be larger than that of stoponium, because of the stronger perturbative color interactions between diquarks. Intuitively, for such a heavy onium, the potentials of both stoponium and diquarkonium should be essentially coulombic and therefore calculable in a model-independent way. However both the potential model estimation \cite{Hagiwara:1990sq} and the recent lattice study \cite{Kim:2015zqa} on the stoponium wave function exhibit substantial departure from coulombic limit. For the case of diquarkonium, only potential model estimation is available \cite{Hagiwara:1990sq}: it was estimated as $\vert R(0) \vert^2/m_{\eta_\Phi}^2 \simeq 1.5$ GeV for a $750$ GeV diquarkonium.

Using the potential model estimation of the wave function at the origin, we then find $\sigma(pp \to \eta_{\Phi^{+4/3}} \to \gamma \gamma)$ to be about $12$ fb for $m_{\eta_\Phi}=750$ GeV at 13~TeV LHC, which may be slightly larger to account for the LHC di-photon excess. Correspondingly, $\sigma(pp \to \eta_\Phi \to gg)$ is predicted to be about $1.6$ pb and is within the experimental bounds \cite{dijetss}. 

Scaling by fourth power of electric charge, one can also obtain 
\bea
\sigma(pp \to \eta_{\Phi^{-2/3}} \to \gamma \gamma)&=&{1\over 16} \sigma(pp \to \eta_{\Phi^{+4/3}} \to \gamma \gamma)\nonumber\\
\sigma(pp \to \eta_{\Phi^{+1/3}} \to \gamma \gamma)&=&{1\over 256} \sigma(pp \to \eta_{\Phi^{+4/3}} \to \gamma \gamma)~.
\eea
The prediction of $\eta_{\Phi^{-2/3}}$ is then too small to account for the LHC di-photon excess~\footnote{There is no lattice study on the diquarkonium system yet, as far as we know. If future lattice study also found the production rate of $\eta_\Phi$ to be $3\sim 4$ times larger than the potential model estimation, just like the case of stoponium reported in \cite{Kim:2015zqa}, the diquarkonium formed by another color-sextet $SU(2)_L$ singlet Higgs $\Phi_6(6,1,-2/3)$ could be a candidate to interpret the LHC di-photon excess with $\sigma(pp \to \eta_\Phi \to \gamma \gamma)$ predicted to be around $3\sim 4$ fb. }.

Another example of exotic onium is color-singlet octetonium formed by a pair of color-octet scalars $\phi_8$ with the quantum numbers $(8, 2, 1/2)$ \cite{Manohar:2006ga}. Such octectonium state has been carefully studied in \cite{Kim:2008bx, Kim:2009mha, Idilbi:2010rs}. The di-photon and digluon decay widths of this octetonium at the leading order can be found as
\bea
\Gamma(\eta_{8} \to \gamma\gamma)&=16 \alpha^2  {\mid R(0)\mid^{2} \over m^{2}_{\eta_8}}~, \\
\Gamma(\eta_{8} \to gg)&=18 \alpha_S^2{\mid R(0)\mid^{2} \over m^{2}_{\eta_8}}~,
\eea
with $\vert R(0) \vert^2/m_{\eta_8}^2 \simeq 1.2$ GeV \cite{Hagiwara:1990sq} for a $750$ GeV octetonium. In addition,  total decay width of octectonium also depends on the Yukawa coupling $\eta_{U}$ due to new decay channel  \cite{Kim:2008bx, Kim:2009mha, Idilbi:2010rs} such as top quark pairs and $\sigma(pp \to \eta_8 \to \gamma \gamma)$ 
can well be of the right magnitude to explain the LHC di-photon excess as in \cite{Kim:2008bx, Kim:2009mha, Idilbi:2010rs} for some choice of $\eta_{U}$.

\section{LHC bounds on Color Exotics without parity}

As a QCD machine, the large production rates of color exotics enable the possibility of early discovery or put stringent bounds on such color exotics based on existing data from LHC or Tevatron.

In the case of supersymmetric theory, if $R$-parity is not broken,
the final state of the cascade decay is always a stable particle with odd $R$-parity.
Constrained by astrophysics and cosmology, such stable particle must be electric neutral
and the color singlet thus can be identified as the dark matter candidate. Such final state then
appears as missing transverse energy $\cancel{E}_{T}$. As we discussed, the pair production of stop in compressed supersymmetry leads to monojet plus $\cancel{E}_{T}$ which does not have much kinematic handle.

On the other hand, without such parity, color exotics as diquark or color octet scalars will then decay
into quark pairs.
If the exotics decay into light jets, pair production of such exotics then leads to four jets final states with two di-jet resonance and
the production rate with 375~GeV resonance is about 36~pb \cite{willie} for color sextet diquark and slightly high for octet states. Since the four jets arise from heavy resonance decay,  the four jets are all of high $p_{T}$ which can significantly reduce the SM four jets background. LHC run-I has excluded coloron octet up to 800~GeV and anti-triplet diquark up to 350~GeV \cite{Khachatryan:2014lpa}.
The sextet diquark production rate is slightly lower than the octet and the 375~GeV diquark decaying into dijet must be excluded already. The diquark with electric charge $-2/3$ that only couples to the down-type quarks and other diquark states that only couple to first two generations fall into this category and have been completely excluded. The diquark with electric charge $+4/3$  can also decay into same-sign top quark pair as $\Phi_{6}\to tt$. The pair production of diquark then leads to four top final state, in particular, with same-sign di-lepton plus jets and $\cancel{E}_{T}$ which fall into the
regular supersymmetric Majorana gluino search  or sgluon search and such light diquark states suffer from severe constraint of gluino bound~\cite{willie}. The sgluon pair with decay into four-top final state has completely excluded 
such possibility~\cite{4top}.

\begin{figure}[htbp]
\begin{center}
\includegraphics[width=6cm, scale=1.0]{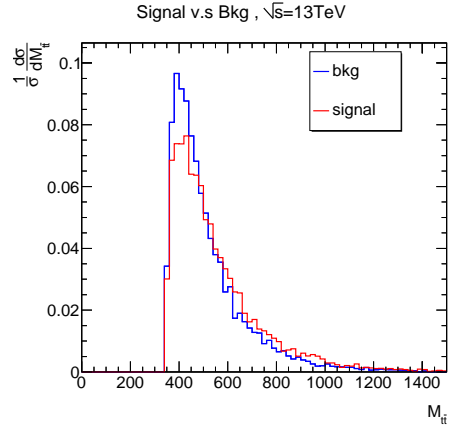}
\caption{ Normalized $t\bar{t}$ invariant mass of SM $pp\to t\bar{t}$ and $pp\to \Phi\Phi^{*}\to t\bar{t}jj$. }
\label{mtt}
\end{center}
\end{figure}

On the other hand, if the decay final state consists of one top quark, the pair production will then fall into $t\bar{t}$ plus jets. For instance, $f_{13}$ dominates the $\Phi^{+4/3}_{6}$ or $f_{33}$ dominates the $\Phi^{+1/3}_{6}$ couplings or charged color octet scalars $\phi^{+}_{8}$ with large top coupling. Then these exotics decay into top
plus light jet as
\beq
\Phi^{+4/3}_{6}\to t+u;~~~~
\Phi^{+1/3}_{6}\to t+b;~~~~
\phi^{+}_{8}\to t+\bar{b}~.
\eeq
We plot the normalized invariant mass distribution $M_{t\bar{t}}$ of SM $pp\to t\bar{t}$ and $pp\to \Phi\Phi^{*}\to t\bar{t}jj$ in Fig. \ref{mtt}.
It is clear that the $t\bar{t}$ invariant mass of 375~GeV resonance decay has the almost identical feature as the SM.

First of all, the diquark production rate of $M_{\Phi}=375~\text{GeV}$ are list as the following
\bea
\sigma(p\bar{p}\to \Phi_{6}\Phi^{*}_{6})_{\sqrt{s} = 1.96~\text{TeV},M_{\Phi}=375~\text{GeV}} &\simeq &  0.01~\text{pb}\nonumber\\
\sigma(pp\to \Phi_{6}\Phi^{*}_{6})_{\sqrt{s}=7~\text{TeV}, M_{\Phi}=375~\text{GeV}} & \simeq &  3.3~\text{pb}\nonumber\\
\sigma(pp\to \Phi_{6}\Phi^{*}_{6})_{\sqrt{s}=14~\text{TeV}, M_{\Phi}=375~\text{GeV}} & \simeq &  36~\text{pb}~.
\eea
In comparison with the $t\bar{t}$ production rates are measured at Tevatron and 7~TeV LHC \cite{tevatrontop,cmstop}
\bea
\sigma(p\bar{p}\to t\bar{t})^{\rm  exp}_{\sqrt{s}=1.96~\text{TeV}} &=& 7.60\pm 0.41({\rm stat})\pm 0.20{\rm syst})\pm 0.36({\rm lumi})~\text{pb}\nonumber\\
\sigma(pp\to t\bar{t})^{\rm  exp}_{\sqrt{s}=7~\text{TeV}}&=& 158\pm 2({\rm stat})\pm 10({\rm syst})\pm 4({\rm lumi})~\text{pb}
\eea
For the exiting data with lower central mass energy collisions, the production rate of diquark are significantly lower than the
systematic errors. Even for 14~TeV LHC, the production is less than 5\% of the SM prediction of $t\bar{t}$ with invariant mass
peak at the SM $t\bar{t}$ threshold. Therefore, we argue the color exotics with decay into one top quark plus a light jet may completely be buried in the SM $t\bar{t}$ samples. 

Such resonance with $t+q$ decay requires $t+j$ reconstruction which has been carried in \cite{tj,Chatrchyan:2013oba}. The resonant $t+j$ reconstruction was designed to search for $W^{\prime}_{R}$ with $g d \to t W^{\prime}$ production and $W^{\prime}_{R}\to \bar{t} b$. For $t+j$ invariant mass $M_{tj}\sim 375$~GeV, the exclusion limit is over 10~pb for 7~TeV LHC by ATLAS. Due to phase-space suppression, both $\Phi_{6}$ and $\phi_{8}$ with resonant mass of 375~GeV are significantly less than that exclusion. The CMS has also excluded spin-1/2 excited quark $t^{'}\to t+g$ in a window between 465 and 512~GeV \cite{Chatrchyan:2013oba}. The production rate for the spin-1/2 exotic quark is about 3 pb for $m_{t^{*}}=375$~GeV at 7~TeV and very close to the production rate of sextet diquark. We then argue the exclusion can be directly applied here and the light diquark decay into
top plus jet can evade the CMS search in \cite{Chatrchyan:2013oba}.

However, color octet scalars $\phi_{8}: (8,2,1/2)$ form a $SU(2)$ doublet with two physical states. In addition to the $\phi^{+}_{8}$ which can dominantly decay into $t\bar{b}$, there always exists a electric neutral state
with degenerate mass. The neutral state $\phi^{0}_{8}\to t\bar{t}$ has been excluded by the four-top sgluon search~\cite{4top} and $\phi^{0}_{8}\to b\bar{b}$ has been excluded by the four light jets search~\cite{Khachatryan:2014lpa}.

On the other hand, the above constraints are all due to the case with prompt decay. Long-lived particle(LLP) searches on the LHC are very dependent on the estimated proper decay length, electric charge and LLP masses. Both the ATLAS and CMS collaborations have published constrains on LLPs with various strategies, including specific energy loss\cite{Aad:2015qfa,Aad:2015oga}, time of flight\cite{ATLAS:2014fka,Chatrchyan:2013oca}, displaced vertex\cite{Aad:2015rba,CMS:2014wda}, HCal detection\cite{Aad:2015asa} and stopped decay\cite{Aad:2013gva,Khachatryan:2015jha}. The very features of LLPs are large ionisation energy loss rate $\frac{dE}{dx}$ and long time of flight. Heavy exotics $\Phi_{6}$ and $\phi_{8}$ are produced with small velocities, some even reaching a non-relativistic region. They hadronize immediately into colorless bound states with different electric charges. The small $\beta\gamma$ values of charged states lead to an anomalously large energy loss rate measured in the tracker. In the meantime, if they have lifetime greater than $\cal O$(ns), they could traverse the whole detector and leave a signal with large time of flight measured in the muon detector. These events are generally selected out with either a muon trigger or a calorimeter-based $\cancel{E}_T$ trigger. In the muon trigger case, the charged colorless bound states would fake heavy muon-like signals in the muon system. Further analysis could incorporate their distinctive energy loss rate for the particle identification. Even if they decay before hitting on the outmost layer, the decay products should not include top quarks as the muon leptons from top decays may still trigger the event selection algorithm. So $\phi_{8}$ which decays to $t\bar{t}$ or $t\bar{b}$ can hardly escape the full-detector search strategy. However, the charged bound states could interact with the detector material and eventually arrive as a neutral state. Then $\cancel{E}_T$ trigger works as a compensation in this case. For example, in the ATLAS $R$-hadron search, gluino mass is excluded to 1270~GeV with the full-detector information and 1260~GeV with the muon-agnostic information \cite{ATLAS:2014fka}.

We assume the $\Phi^{+4/3}_{6}$ is metastable and focus on its searches in the following. Metastable sextet diquarks form neutral, singly and doubly charged hadrons like $\Phi_{6}\bar{u}\bar{u}$, $\Phi_{6}d$ and $\Phi_{6}u$ soon after their production and decay as, for instance, 
\beq
X^{++}(\Phi_{6}u)\to p^{+}+\pi^{+}~.
\eeq
Displaced decay searches are sensitive to multi-jet signals in the tracker volume. The CMS tracker is able to detect a long-lived neutral particle X in a mass range of 50~GeV to 350~GeV decaying to $q\bar q$ pair\cite{CMS:2014wda}. The production cross-section limit depends on the mean proper decay length of X, but always below that of $\Phi_{6}$ pair production. As the upper exclusion mass limit is very close to 375~GeV, we assume technically it can be extended to probe such mass region. In the scenario where the LLPs could traverse the tracker and then decay into light jet pairs in the calorimeters, more calorimeter-based strategies are effective. Though the previous jet pair resonance search doesn't apply here because of the strict reconstructed primary vertex requirement for a four jets event\cite{Khachatryan:2014lpa}, such high $p_T$ jet events are recorded and it's easy to check whether a resonance could be reconstructed. Moreover, the ATLAS calorimeter signal searches covers a radius from the outer ECal to the HCal and looks for metastable neutral particles that decay in this region\cite{Aad:2015asa}. The production cross section limit covers a mass region up to 150~GeV for a pair of such neutral states decaying from a heavy hidden scalar. The CMS displaced decay and ATLAS HCal decay searches are designed for neutral LLPs, but we expect the charged hadronic states that would inevitably leave significant charged tracks are detectable within these methods which has never been reported. 
At last, there's a scenario where bound states with small $\beta$ values may stop in the detectable region due to the energy loss when interacting with detector material. Then there should be an upper lifetime limit set by the out-of-time decay searches where none active event is expected in the detector\cite{Aad:2013gva,Khachatryan:2015jha}. In general, one expects much more severe bound over the charged LLP in comparison with neutral LLP which is already excluded up to 1.2 TeV for octet \cite{ATLAS:2014fka}.

\section{Conclusions}

In summary, we discuss the interpretation of the recent di-photon excess as onia of massive meta-stable colored and charged particles. To escape from the stringent bound on di-lepton final states from vector onia decays, the onia should be formed by scalar particles. One example is stoponium which has been discussed in details in \cite{martin,Kim:2014yaa}, however the predicted di-photon signal is significantly smaller than the required rate to explain the LHC di-photon excess. We then consider various color exotics as sextet diquark or color octet scalar. Diphoton from Octetonium can well be of the right magnitude to explain the LHC di-photon excess as in \cite{Kim:2008bx, Kim:2009mha, Idilbi:2010rs} for some parameter choice. Diquarkonium formed by color sextet diquark with electric charge to be $4/3$ also predicts the di-photon signal as 12 fb at 13 TeV LHC  with the enhancement from color factor and electric charge. On the other hand, light color exotics suffer from severe constraints of direct searches at the hadron colliders. Color sextet or octet scalars with decay into light jets or top pairs have been excluded at this mass range. We
find the only viable channel is when $\Phi^{+4/3}\to u+t$ dominates the diquark decay which predicts resonance with top plus jet and is still below the current bound.

\section*{Acknowledgement}

The work is supported in part by the National Science Foundation of China (11135006,  11275168, 11422544, 11075139, 11375151, 11535002) and the Zhejiang University Fundamental Research Funds for the Central Universities. KW is also supported by Zhejiang University K.P.Chao High Technology Development Foundation.
GZ is also supported by the Program for New Century Excellent
Talents in University (Grant No. NCET-12-0480).


\begin{thebibliography}{99}


\bibitem{atlas}
  The ATLAS collaboration,
  ATLAS-CONF-2015-081.

\bibitem{CMS:2015dxe}
  CMS Collaboration [CMS Collaboration],
  CMS-PAS-EXO-15-004.


\bibitem{Randall:2008xg} 
  L.~Randall and M.~B.~Wise,
  arXiv:0807.1746 [hep-ph].


\bibitem{Aad:2015kna} 
  G.~Aad {\it et al.} [ATLAS Collaboration],
  Eur.\ Phys.\ J.\ C {\bf 76}, no. 1, 45 (2016)
  doi:10.1140/epjc/s10052-015-3820-z
  [arXiv:1507.05930 [hep-ex]].
  G.~Aad {\it et al.} [ATLAS Collaboration],
  JHEP {\bf 1601}, 032 (2016)
  doi:10.1007/JHEP01(2016)032
  [arXiv:1509.00389 [hep-ex]].
  V.~Khachatryan {\it et al.} [CMS Collaboration],
  JHEP {\bf 1510}, 144 (2015)
  doi:10.1007/JHEP10(2015)144
  [arXiv:1504.00936 [hep-ex]].
  
  
\bibitem{Backovic:2015fnp}
  M.~Backovic, A.~Mariotti and D.~Redigolo,
  arXiv:1512.04917 [hep-ph].
  K.~Harigaya and Y.~Nomura,
  arXiv:1512.04850 [hep-ph].
  Y.~Mambrini, G.~Arcadi and A.~Djouadi,
  arXiv:1512.04913
  A.~Angelescu, A.~Djouadi and G.~Moreau,
  arXiv:1512.04921 [hep-ph].
  A.~Pilaftsis,
  arXiv:1512.04931 [hep-ph].
  R.~Franceschini {\it et al.},
  arXiv:1512.04933 [hep-ph].
  S.~Di Chiara, L.~Marzola and M.~Raidal,
  arXiv:1512.04939 [hep-ph].
  M.~Low, A.~Tesi and L.~T.~Wang,
  arXiv:1512.05328 [hep-ph].
  B.~Bellazzini, R.~Franceschini, F.~Sala and J.~Serra,
  arXiv:1512.05330 [hep-ph].
  J.~Ellis, S.~A.~R.~Ellis, J.~Quevillon, V.~Sanz and T.~You,
  arXiv:1512.05327 [hep-ph].
  S.~D.~McDermott, P.~Meade and H.~Ramani,
  arXiv:1512.05326 [hep-ph].
  T.~Higaki, K.~S.~Jeong, N.~Kitajima and F.~Takahashi,
  arXiv:1512.05295 [hep-ph].
  R.~S.~Gupta, S.~JÃ€ger, Y.~Kats, G.~Perez and E.~Stamou,
  arXiv:1512.05332 [hep-ph].
  C.~Petersson and R.~Torre,
  arXiv:1512.05333 [hep-ph].
  E.~Molinaro, F.~Sannino and N.~Vignaroli,
  arXiv:1512.05334 [hep-ph].
  Y.~Nakai, R.~Sato and K.~Tobioka,
  arXiv:1512.04924 [hep-ph].
  D.~Buttazzo, A.~Greljo and D.~Marzocca,
  arXiv:1512.04929 [hep-ph].
  Y.~Bai, J.~Berger and R.~Lu,
  arXiv:1512.05779 [hep-ph].
  D.~Aloni, K.~Blum, A.~Dery, A.~Efrati and Y.~Nir,
  arXiv:1512.05778 [hep-ph].
  A.~Falkowski, O.~Slone and T.~Volansky,
  arXiv:1512.05777 [hep-ph].
  C.~Csaki, J.~Hubisz and J.~Terning,
  arXiv:1512.05776 [hep-ph].
  A.~Ahmed, B.~M.~Dillon, B.~Grzadkowski, J.~F.~Gunion and Y.~Jiang,
  arXiv:1512.05771 [hep-ph].
  J.~Chakrabortty, A.~Choudhury, P.~Ghosh, S.~Mondal and T.~Srivastava,
  arXiv:1512.05767 [hep-ph].
  L.~Bian, N.~Chen, D.~Liu and J.~Shu,
  arXiv:1512.05759 [hep-ph].
  D.~Curtin and C.~B.~Verhaaren,
  arXiv:1512.05753 [hep-ph].
  S.~Fichet, G.~von Gersdorff and C.~Royon,
  arXiv:1512.05751 [hep-ph].
  W.~Chao, R.~Huo and J.~H.~Yu,
  arXiv:1512.05738 [hep-ph].
  S.~V.~Demidov and D.~S.~Gorbunov,
  arXiv:1512.05723 [hep-ph].
  J.~M.~No, V.~Sanz and J.~Setford,
  arXiv:1512.05700 [hep-ph].
  D.~Becirevic, E.~Bertuzzo, O.~Sumensari and R.~Z.~Funchal,
  arXiv:1512.05623 [hep-ph].
  P.~Cox, A.~D.~Medina, T.~S.~Ray and A.~Spray,
  arXiv:1512.05618 [hep-ph].
  R.~Martinez, F.~Ochoa and C.~F.~Sierra,
  arXiv:1512.05617 [hep-ph].
  A.~Kobakhidze, F.~Wang, L.~Wu, J.~M.~Yang and M.~Zhang,
  arXiv:1512.05585 [hep-ph].
  S.~Matsuzaki and K.~Yamawaki,
  arXiv:1512.05564 [hep-ph].
  Q.~H.~Cao, Y.~Liu, K.~P.~Xie, B.~Yan and D.~M.~Zhang,
  arXiv:1512.05542 [hep-ph].
  B.~Dutta, Y.~Gao, T.~Ghosh, I.~Gogoladze and T.~Li,
  arXiv:1512.05439 [hep-ph].
  R.~Benbrik, C.~H.~Chen and T.~Nomura,
  arXiv:1512.06028 [hep-ph].
  E.~Megias, O.~Pujolas and M.~Quiros,
  arXiv:1512.06106 [hep-ph].
  J.~Bernon and C.~Smith,
  arXiv:1512.06113 [hep-ph].
  L.~M.~Carpenter, R.~Colburn and J.~Goodman,
  arXiv:1512.06107 [hep-ph].
  A.~Alves, A.~G.~Dias and K.~Sinha,
  arXiv:1512.06091 [hep-ph].
  E.~Gabrielli, K.~Kannike, B.~Mele, M.~Raidal, C.~Spethmann and H.~VeermÃ€e,
  arXiv:1512.05961 [hep-ph].
\bibitem{Agrawal:2015dbf}
  P.~Agrawal, J.~Fan, B.~Heidenreich, M.~Reece and M.~Strassler,
  arXiv:1512.05775 [hep-ph].
  

\bibitem{nojiri}
  M.~Drees and M.~M.~Nojiri,
  Phys.\ Rev.\ D {\bf 49}, 4595 (1994)
  doi:10.1103/PhysRevD.49.4595
  [hep-ph/9312213].
\bibitem{martin}
  S.~P.~Martin,
  Phys.\ Rev.\ D {\bf 77}, 075002 (2008)
  doi:10.1103/PhysRevD.77.075002
  [arXiv:0801.0237 [hep-ph]].

\bibitem{Kim:2014yaa} 
  C.~Kim, A.~Idilbi, T.~Mehen and Y.~W.~Yoon,
  Phys.\ Rev.\ D {\bf 89}, no. 7, 075010 (2014)
  doi:10.1103/PhysRevD.89.075010
  [arXiv:1401.1284 [hep-ph]].
  
\bibitem{Kim:2015zqa}
  S.~Kim,
  Phys.\ Rev.\ D {\bf 92}, no. 9, 094505 (2015)
  doi:10.1103/PhysRevD.92.094505
  [arXiv:1508.07080 [hep-lat]].
  
\bibitem{Hagiwara:1990sq}
  K.~Hagiwara, K.~Kato, A.~D.~Martin and C.~K.~Ng,
  Nucl.\ Phys.\ B {\bf 344}, 1 (1990).
  doi:10.1016/0550-3213(90)90683-5

  
\bibitem{Manohar:2006ga} 
  A.~V.~Manohar and M.~B.~Wise,
  Phys.\ Rev.\ D {\bf 74}, 035009 (2006)
  doi:10.1103/PhysRevD.74.035009
  [hep-ph/0606172].


\bibitem{Chacko:1998td}
  Z.~Chacko and R.~N.~Mohapatra,
  Phys.\ Rev.\  D {\bf 59}, 055004 (1999)
  [arXiv:hep-ph/9802388].
    R.~N.~Mohapatra, N.~Okada and H.~B.~Yu,
  Phys.\ Rev.\  D {\bf 77}, 011701 (2008)
  [arXiv:0709.1486 [hep-ph]].


\bibitem{willie}
  C.~R.~Chen, W.~Klemm, V.~Rentala and K.~Wang,
  Phys.\ Rev.\ D {\bf 79}, 054002 (2009)
  doi:10.1103/PhysRevD.79.054002
  [arXiv:0811.2105 [hep-ph]].



\bibitem{babu}
  K.~S.~Babu, R.~N.~Mohapatra and S.~Nasri,
  Phys.\ Rev.\ Lett.\  {\bf 97}, 131301 (2006)
  [arXiv:hep-ph/0606144].



\bibitem{chen}
  C.~H.~Chen,
  Phys.\ Lett.\ B {\bf 680}, 133 (2009)
  doi:10.1016/j.physletb.2009.08.045
  [arXiv:0902.2620 [hep-ph]].

  
\bibitem{dijetss}
  [ATLAS Collaboration],
  arXiv:1512.01530 [hep-ex].
  V.~Khachatryan {\it et al.} [CMS Collaboration],
  arXiv:1512.01224 [hep-ex].
  
\bibitem{Kim:2008bx} 
  C.~Kim and T.~Mehen,
  Phys.\ Rev.\ D {\bf 79}, 035011 (2009)
  doi:10.1103/PhysRevD.79.035011
  [arXiv:0812.0307 [hep-ph]].

\bibitem{Kim:2009mha} 
  C.~Kim and T.~Mehen,
  Nucl.\ Phys.\ Proc.\ Suppl.\  {\bf 200-202}, 179 (2010)
  doi:10.1016/j.nuclphysbps.2010.02.081
  [arXiv:0909.5695 [hep-ph]].
  
\bibitem{Idilbi:2010rs} 
  A.~Idilbi, C.~Kim and T.~Mehen,
  Phys.\ Rev.\ D {\bf 82}, 075017 (2010)
  doi:10.1103/PhysRevD.82.075017
  [arXiv:1007.0865 [hep-ph]].

\bibitem{Khachatryan:2014lpa} 
  V.~Khachatryan {\it et al.} [CMS Collaboration],
  Phys.\ Lett.\ B {\bf 747}, 98 (2015)
  doi:10.1016/j.physletb.2015.04.045
  [arXiv:1412.7706 [hep-ex]].
  



\bibitem{4top} 
  G.~Aad {\it et al.} [ATLAS Collaboration],
  JHEP {\bf 1508}, 105 (2015)
  doi:10.1007/JHEP08(2015)105
  [arXiv:1505.04306 [hep-ex]].


\bibitem{tevatrontop}
  R.~Y.~Peters [CDF and D0 Collaborations],
  arXiv:1509.07629 [hep-ex].

\bibitem{cmstop}
  S.~Tosi [CMS Collaboration],
  EPJ Web Conf.\  {\bf 95}, 04069 (2015).
  doi:10.1051/epjconf/20159504069

\bibitem{tj} 
  G.~Aad {\it et al.} [ATLAS Collaboration],
  Phys.\ Rev.\ D {\bf 86}, 091103 (2012)
  doi:10.1103/PhysRevD.86.091103
  [arXiv:1209.6593 [hep-ex]].
\bibitem{Chatrchyan:2013oba} 
  S.~Chatrchyan {\it et al.} [CMS Collaboration],
  JHEP {\bf 1406}, 125 (2014)
  doi:10.1007/JHEP06(2014)125
  [arXiv:1311.5357 [hep-ex]].
  
  
  
\bibitem{Aad:2015qfa} 
  G.~Aad {\it et al.} [ATLAS Collaboration],
  Eur.\ Phys.\ J.\ C {\bf 75}, no. 9, 407 (2015)
  doi:10.1140/epjc/s10052-015-3609-0
  [arXiv:1506.05332 [hep-ex]].
  
\bibitem{Aad:2015oga} 
  G.~Aad {\it et al.} [ATLAS Collaboration],
  Eur.\ Phys.\ J.\ C {\bf 75}, 362 (2015)
  doi:10.1140/epjc/s10052-015-3534-2
  [arXiv:1504.04188 [hep-ex]].
  
\bibitem{ATLAS:2014fka} 
  G.~Aad {\it et al.} [ATLAS Collaboration],
  JHEP {\bf 1501}, 068 (2015)
  doi:10.1007/JHEP01(2015)068
  [arXiv:1411.6795 [hep-ex]].
  
\bibitem{Chatrchyan:2013oca} 
  S.~Chatrchyan {\it et al.} [CMS Collaboration],
  JHEP {\bf 1307}, 122 (2013)
  doi:10.1007/JHEP07(2013)122
  [arXiv:1305.0491 [hep-ex]].

\bibitem{Aad:2015rba} 
  G.~Aad {\it et al.} [ATLAS Collaboration],
  Phys.\ Rev.\ D {\bf 92}, no. 7, 072004 (2015)
  doi:10.1103/PhysRevD.92.072004
  [arXiv:1504.05162 [hep-ex]].

\bibitem{CMS:2014wda} 
  V.~Khachatryan {\it et al.} [CMS Collaboration],
  Phys.\ Rev.\ D {\bf 91}, no. 1, 012007 (2015)
  doi:10.1103/PhysRevD.91.012007
  [arXiv:1411.6530 [hep-ex]].
  
  
\bibitem{Aad:2015asa} 
  G.~Aad {\it et al.} [ATLAS Collaboration],
  Phys.\ Lett.\ B {\bf 743}, 15 (2015)
  doi:10.1016/j.physletb.2015.02.015
  [arXiv:1501.04020 [hep-ex]].
  
  
\bibitem{Aad:2013gva} 
  G.~Aad {\it et al.} [ATLAS Collaboration],
  Phys.\ Rev.\ D {\bf 88}, no. 11, 112003 (2013)
  doi:10.1103/PhysRevD.88.112003
  [arXiv:1310.6584 [hep-ex]].

\bibitem{Khachatryan:2015jha} 
  V.~Khachatryan {\it et al.} [CMS Collaboration],
  Eur.\ Phys.\ J.\ C {\bf 75}, no. 4, 151 (2015)
  doi:10.1140/epjc/s10052-015-3367-z
  [arXiv:1501.05603 [hep-ex]].


\end{thebibliography}
\end{document}